\documentclass[letterpaper, 10 pt, conference]{ieeeconf}

\IEEEoverridecommandlockouts                              
\usepackage{graphicx}
\graphicspath{{figure/}}
\usepackage{graphics} 
\usepackage{times} 
\usepackage{amssymb}
\usepackage[colorlinks=false, urlcolor=blue, pdfborder={0 0 0}]{hyperref}
\usepackage{enumerate} 
\usepackage{subfigure}
\usepackage{comment}
\usepackage{mathrsfs}
\usepackage{subfigure}
\usepackage{url}
\usepackage{wrapfig}
\usepackage[ruled,vlined]{algorithm2e}
\usepackage{amsmath}
\usepackage{array}
\usepackage{dsfont}
\usepackage{amsfonts}
\usepackage{color}
\usepackage{amssymb}
\usepackage{multicol}
\usepackage{cite}
\usepackage{mathrsfs}
\usepackage[T1]{fontenc}

\newtheorem{remark}{Remark}
\newtheorem{example}{Example}

\title
{\LARGE \bf Modeling and Analysis of Networked  Discrete Event Systems \\ with Multiple Control Channels}

\author{Zhaocong Liu, Junyao Hou, Xiang Yin  and Shaoyuan Li
\thanks{This work was  supported by the National Natural Science Foundation of China (61803259, 61833012) and by Shanghai Jiao Tong University Scientific and Technological Innovation Funds.}
\thanks{Zhaocong Liu, Junyao Hou, Xiang Yin and Shaoyuan Li are with Department of Automation and Key Laboratory of System Control and Information Processing, Shanghai Jiao Tong University, Shanghai 200240, China
	E-mail: {\tt\small  \{zhaocongl,houjunyao,yinxiang\}@sjtu.edu.cn}.}
}
 
\newcommand{\Sup}{\textbf{\textsc{Sup}}}
\newcommand{\Channel}{\textbf{\textsc{Channel}}}
\newcommand{\Act}{\textbf{\textsc{Act}}}
\newcommand{\1}{\textbf{1}}
	\newcommand{\0}{\textbf{0}}
\begin{document}

\maketitle

\begin{abstract}
In this paper, we propose a novel framework for modeling and analysis of networked discrete-event systems (DES). 
We assume that the plant is controlled by a feedback supervisor whose control decisions are subject to communication delays and losses. 
Furthermore, we consider a general setting where the supervisor sends control decisions to different actuators via different communication channels whose dynamics are independent.   
We provide a system theoretic approach by  identifying the state-space of overall networked system and investigating the dynamic of the entire state-space.  
Our approach precisely specifies the roles of the supervisor, the communication channels and the actuators.  
Also, we compare the proposed networked DES model with the existing one and show that the proposed networked model captures  physical situations of  networked systems more precisely.
\end{abstract}

\section{Introduction}
Supervisory control theory (SCT) in the context of Discrete Event Systems (DESs) is a formal approach for the synthesis of correct-by-construction controllers for dynamic systems. 
The SCT has been extensively investigated for more than thirty years since the seminal work of Ramadge and Wonham \cite{wonham2018supervisory}. 
Nowadays, in many modern applications, supervisors are connected with plants via \emph{communication networks}   to transmit data, e.g., control decisions or sensor readings. This leads to the so called \emph{networked discrete event systems}, which have drawn considerable attention in the past few years; see, e.g., \cite{lin2014control,nunes2018codiagnosability,rashidinejad2018supervisory,cdc19,zhu2019supervisor,alves2019state,sasi2018detectability,lin2020information,hou2019relative,lin2019state}.

Compared with non-networked control architecture, one of the most important features of networked control architecture is that the control decision module, e.g., the supervisor, may not be embedded in the plant but are connected with plants via networks. In many applications, for example, this is due to physical constraints because sensors, actuators and supervisors are physically distributed, in particular, for large-scale  systems. 
Moreover, the networked control architecture provides more flexibility for the implementation of the supervisors. For example, one can implement the supervisor in the cloud by utilizing more powerful computational resources. 

On the other hand, transmitting data over networks is sometimes unreliable due to communication delays and losses; this is particularly the case in  process industries with severe operation environments \cite{lunze2014control}. 
Therefore, the central problem in networked DES is  to appropriately handle  communication delays and losses in the control/observation channels, which is an active research direction in the literature.  
For example, in \cite{balemi1994input,zgorzelski2018new,zgorzelski2019model}, I/O automata are used to model networked DES with communication delays.  
In \cite{park2006delay,zhang2016delay}, the robustness of supervisors with networks has been investigated. Timing issue has also been considered in networked DES by \cite{rashidinejad2018supervisory,pruekprasert2019supervisory,zhao2017supervisory,zhang2016delay,alves2017supervisory}.

In the recent work of Lin \cite{lin2014control} and its subsequential works \cite{shu2014decentralized,komenda2016modular,shu2017predictive,shu2017deterministic}, a language-based framework is proposed for  modeling and control of networked DES. In particular, communication delays and losses are considered in both the control channel and the observation channel. In this framework, necessary and sufficient conditions for the existence of a supervisor as well as supervisor synthesis procedures have been investigated extensively. 
Our work is motivated by the language-based framework proposed by Lin. 
However, we aim to further explore more practical networked DES models by considering the following two issues that have not been fully addressed in \cite{lin2014control}:
\begin{itemize}
	\item 
	The control decision in the framework of \cite{lin2014control} is sent as a ``package" to all actuators; this corresponds to the case of a \emph{single} control channel. 
	In practice, actuators in the plant may be physically distributed and, therefore, different actuators, which correspond to different controllable events, may receive decisions from the supervisor by different control channels.
	\item 
	The closed-loop language defined in \cite{lin2014control} is an \emph{over-approximation} of the exact language of the networked systems. Particularly, there may exist physically infeasible strings in the closed-loop language defined in  \cite{lin2014control}; 
	an example for such a scenario is provided in Section~\ref{sec:5}. 
\end{itemize}

\begin{figure}
	\centering
	\includegraphics[scale=0.45]{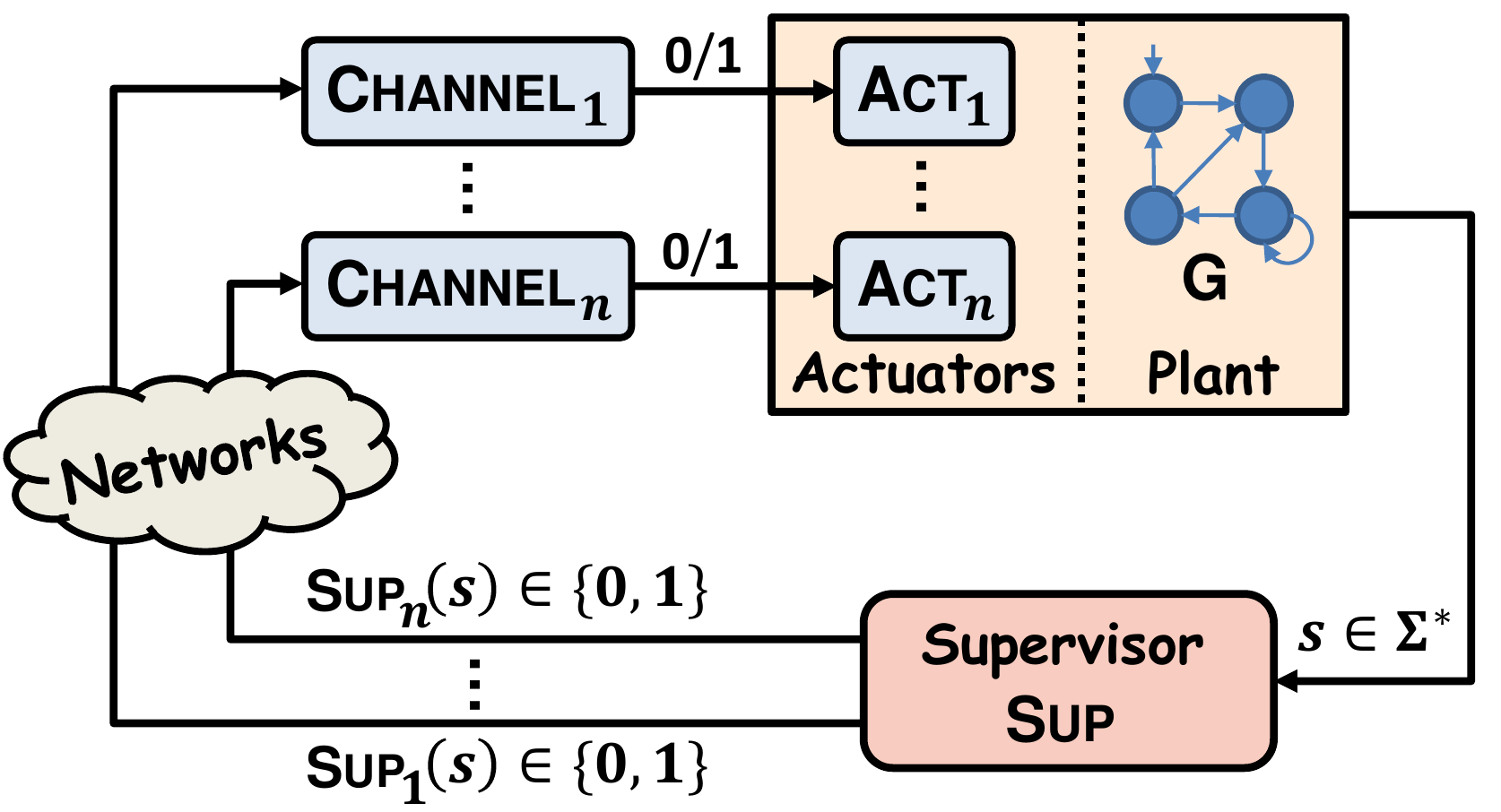}
	\caption{Networked supervisory control architecture with multiple control channels considered in this work.}\label{fig:diag} 
\end{figure} 

In this work, we investigate the networked supervisory control architecture  depicted in Figure~\ref{fig:diag}, where the supervisor sends control decisions to different actuators via different control channels whose dynamics are independent and all channels may be subject to communication delays and losses.  
To focus on investigating the effect of delays and losses in multiple control channels, we assume that the observation channel is perfect with full observation.  
Our approach is from a system theoretic perspective by identifying the state-space of overall networked system and investigating the dynamic of the entire state-space.  
Such ``states" of the entire networked systems are referred to as \emph{network configurations}. 
In particular, we explicitly distinguish  the roles of the supervisor, the communication channels and the actuators because the current control decision issued by the supervisor may not be the current control decision applied in the actuator. 
In particular, we show that, compared with the closed-loop language defined in \cite{lin2014control}, our definition of closed-loop language is more precise in the sense that it excludes those strings that are physically infeasible.

Our work is most related to the recent work of \cite{zhu2019supervisor}, where FIFO channel automata are proposed to model the dynamic of the communication channels. However, the model in \cite{zhu2019supervisor} is for the case of single control channel, while our new dynamic model  captures the nature of multiple control channels. 
In the literature of networked DES, the effect of multiple observation channels has been investigated very recently for the purpose of state estimation and detection \cite{lin2019state,alves2019state,yin2019opacity}.
However,  to the best of our knowledge,  the effect of multiple control channels in the context of networked DES has never been considered. 
Furthermore, we explicitly compare  the closed-loop language in our framework with the one in \cite{lin2014control}, which is also not considered in \cite{zhu2019supervisor}. 

The remainder of this paper is organized as follows. 
In Section~\ref{sec:2}, we introduce some necessary notations and present multi-channel setting. Modeling and analysis of communication delays and communication losses in each single control channel are provided in Section~\ref{sec:3}.
In Section~\ref{sec:4}, we propose the closed-loop dynamics of the overall system consisting of multiple control channels. 
The comparison between our framework and the framework in \cite{lin2014control} is provided in Section~\ref{sec:5}.

\section{Networked Supervisory Control}\label{sec:2}
\subsection{Standard Supervisory Control}
We consider a DES modeled  as a deterministic finite-state automaton (DFA) 
\[
\mathbf{G}=(Q,\Sigma,\delta,q_0),
\]
where $Q$ is the finite set of states, $\Sigma$ is the finite set of events, 
$\delta: Q \times \Sigma \to Q$ is the (partial) transition function and  
$q_0 \in Q$ is the initial state. 
The transition function  can be extended to $\delta: Q \times \Sigma^* \to Q$ recursively in the usual manner; see, e.g., \cite{Lbook}.
A string is a sequence of events in the form of $s=\sigma_1\sigma_2\cdots\sigma_n, \sigma_i\in \Sigma$; a language is a set of strings. 
We denote by $\Sigma^*$ the set of all  strings  over  $\Sigma$ including the empty string $\epsilon$.
Then the language generated by DFA $\mathbf{G}$ is $\mathcal{L} (\mathbf{G}) = \{s\in \Sigma^*: \delta(q_0, s)! \}$, where $!$  means ``is defined".
For any positive natural number $N\in \mathbb{N}$, we denote by $[1, N]$ the set of natural numbers from $1$ to $N$.

In the supervisory control framework, the event set is partitioned as two disjoint sets 
\[
\Sigma = \Sigma_c \dot{\cup} \Sigma_{uc}, 
\]
where $\Sigma_c$ is the set of controllable events and $\Sigma_{uc}$ is the set of uncontrollable events. 
A supervisor is a function 
\begin{equation}\label{eq:sup-stand}
    \Sup:  \mathcal{L} (G) \to 2^{\Sigma_c}, 
\end{equation}
that  dynamically decides which controllable events to enable. 
Note that uncontrollable events are always enabled by default. 
Then the closed-loop language under control, denoted by $\mathcal{L}(\Sup/\mathbf{G})$, is defined recursively by 
\begin{itemize}
	\item 
	$\epsilon\in \mathcal{L}(\Sup/\mathbf{G})$; 
	\item 
	for any $s\in \Sigma^*,\sigma\in \Sigma$, we have $s\sigma \in \mathcal{L}(\Sup/\mathbf{G})$ iff
	\[
	[s\in \mathcal{L}(\Sup/\mathbf{G})]
	\wedge 
	[s\sigma\in \mathcal{L}(\mathbf{G})]
	\wedge 
	[\sigma \in \Sup(s)\cup \Sigma_{uc}].
	\]
\end{itemize}
Intuitively, the above definition says that $s\sigma$ is in the closed-loop language if 
(i) $s$ is in the closed-loop language; and 
(ii) $s\sigma$ is feasible in $\mathbf{G}$; and 
(iii) $\sigma$ is either uncontrollable or enabled by $\Sup$ upon the occurrence of $s$.

\subsection{Supervisor Implementation  over Networks}
Note that we can write supervisor $\Sup$ in Equation~\eqref{eq:sup-stand} equivalently as 
\begin{equation}\label{eq:sup-net}
\Sup:\mathcal{L}(\mathbf{G})\times \Sigma_c\to \{\0,\1\},  
\end{equation}
where ``\0" means ``disable" and ``\1" means ``enable".

Although supervisors defined in Equations~\eqref{eq:sup-stand} and~\eqref{eq:sup-net} are mathematically equivalent, their physical interpretations in terms of how the supervisor is implemented  are quite different. 
Specifically, Equation~\eqref{eq:sup-stand} suggests that, at each instant, the supervisor sends  control decision $\Sup(s)$ \emph{as a package} to the plant. 
However, in practice, each controllable event   represents a single actuator and actuators in the plant may be physically distributed. 
Therefore, Equation~\eqref{eq:sup-net} captures a more general scenario where the supervisor sends control decision  $\Sup(s)(\sigma)$ (disable or enable) to each $\sigma$ \emph{individually}. 
This difference is  crucial in the networked setting as the supervisor may send control decisions to different actuators by different communication channels whose network properties (delays or losses) are different and independent.

In general, the system has $|\Sigma_c|$ actuators.
Hereafter, for the sake of simplicity, we assume that event set $\Sigma$ is ordered so that the first $|\Sigma_c|$ events in $\Sigma$ are controllable with $\Sigma_c = \{\sigma_1, \sigma_2, \cdots, \sigma_{|\Sigma_c|}\}$.
For each $\sigma_i\in \Sigma_c$, the corresponding actuator is denoted by $\Act_i$. 
Also, for the sake of simplicity, we write
\[
\Sup_i(s):=\Sup(s)(\sigma_i), 
\]
which is the control decision (\0 or \1) the supervisor sends to $\Act_i$ upon the occurrence of string $s$ in the plant.
In the non-networked setting, this decision will be received by $\Act_i$ immediately and take effect. 
However, in the networked setting, the decision for $\sigma_i$ is transmitted to  $\Act_i$ via the corresponding control channel denoted by $\Channel_i$, which may be subject to the following issues:
\begin{itemize}
	\item 
	\emph{Control Delays}:
	decision $\Sup_i(s)$ may be delayed in the sense that it will go to $\Channel_i$ and is not received by  $\Act_i$ immediately but after some delays. 
	\item 
	\emph{Control Losses}: 
	decision $\Sup_i(s)$ may be lost in the sense that it will not go to $\Channel_i$, and therefore, it will never be received by $\Act_i$. 
\end{itemize}

Note that, in this work, we only focus on the issue of delays and losses in control channels. The observation channels are assumed to be perfect with  full observation.  
To present our  framework, we make the following assumptions on the control channels:
\begin{enumerate}[{A}1]
	\item
	For each $\Channel_i$, the number of consecutive decision losses is upper bounded by integer $N_i^L\in \mathbb{N}$; 
	\item 
    For each $\Channel_i$, the number of control delays is  upper bounded by integer  $N_i^D\in \mathbb{N}$;
    \item 
    The number of delays or losses are counted by event occurrences in the plant;
    \item 
    For each $\Channel_i$, decisions in the channel are first-in-first-out (FIFO);
    \item 
    The initial control decisions are embedded in each $\Act_i$; hence are not subject to control delays or  losses;
    \item 
    The dynamic of each $\Channel_i, i=1,\dots,|\Sigma_c|$ are independent, i.e., 
    delays or losses in one channel will not affect delays or losses in another channel.
\end{enumerate} 
In the following sections, we will formally describe the dynamics of such a networked supervisory control system. 

\section{Delays and Losses in  Single Channel}\label{sec:3}
Since  control channels are independent, 
in this section, we focus on   modeling and analysis of each \emph{single channel}. 
Specifically, we provide a framework to explicitly capture the issues of delays and losses from a system theoretic perspective. 
 
\subsection{State Space for  Control Networks}

In the system theory,   \emph{states} are  referred to as variables  with  minimum number that are sufficient enough to  represent the current status of the dynamic system. 
In the standard supervisory control framework, 
string $s\in \mathcal{L}(\Sup/\mathbf{G})$ executed is sufficient enough to serve as the ``state" of the system as it determines both the current-state $\delta(q_0,s)$ in the plant and the current control decision $\Sup(s)$ applied (when $\Sup$ is realized by a finite-state automaton, it further suffices to know the current-state in the supervisor automaton). 
However, this information is not sufficient to determine the current status of a networked control system as the current control decision issued  and the current control decision applied may be different due to communication delays and losses.

Now, we provide a general state-space model for networked supervisory control systems. 
The ``state" of the networked closed-loop system contains four components:
\begin{enumerate}[(i)]
	\item 
	The states of the plant and the supervisor; both can be determined by string $s\in \mathcal{L}(\mathbf{G})$ executed. 
	\item 
	The configuration of each actuator $\Act_i$ representing the \emph{current effective control decision} (\0 or \1) for $\sigma_i$.
	\item 
	The configuration of each control channel $\Channel_i$, which captures
	\emph{what decisions are waiting in the channel and in what order}. 
	Formally, for each $\Channel_i$, the channel configuration for $\sigma_i$ is a set of pairs in the form of 
	\[
	\theta_i=\{ (\gamma_1,n_1 ),(\gamma_2,n_2 ),\dots, (\gamma_k,n_k )   \} 
	\]   
	where $\gamma_j\in \{\0,\1\}$ and $n_j\in [1,N_i^D]$. 
	We denote by $\Theta_i$   the set of  all channel configurations for $\Channel_i$. 
	Intuitively, for each $(\gamma,n)\in \theta_i$, $\gamma\in \{\0,\1\}$ represents a disable or enable decision while $n$ represents the \emph{maximum remaining time} of $\gamma$ in the channel. 
	\item 
	Finally, for the case of decision losses, we also need a counter to capture the number of consecutive decision losses for each channel.  
\end{enumerate}

Hereafter, we will discuss how this state-space evolves, i.e., the dynamics of the networked control system. 
We first discuss the issues of communication delays and losses separately and then show how to combine them together.

\subsection{Communication Delays}

For the case of control delays, in addition to the current string executed $s\in \mathcal{L}(\mathbf{G})$, we also need to track the current channel configuration $\theta_i\in \Theta_i$ and the current actuator configuration $a_i\in \{\0,\1\}$. 
For any $\theta_i\in \Theta_i$, we define operator $\textsc{NX}:\Theta_i\to \Theta_i$ by 
\[
\textsc{NX}(\theta_i) 
= \{ (\gamma, n - 1) : (\gamma, n)\in \theta_i \wedge n > 1 \} 
\]
which decreases the remaining time of each control decision in the channel by one unit.

Now, for controllable event $\sigma_i\in \Sigma_c$, 
suppose that the current actuator configuration for $\Act_i$ is $a_i\in \{\0,\1\}$ and the current channel configuration $\Channel_i$ is $\theta_i= \{(\gamma_1,n_1 ),(\gamma_2,n_2 ),\dots, (\gamma_k,n_k ) \} \in \Theta_i$. 
Suppose that $\Sup_i$ sends a new decision $\gamma \in \{\0,\1\}$. 
We are interested in what are the next actuator configuration and channel configuration for $\sigma_i$. 
To this end, we define operator
\[
\texttt{Move}_i^D : \{\0,\1\}\times \Theta_i \times \{\0,\1\} \to 2^{ \{\0,\1\}\times \Theta_i}
\]
by: 
for any  $(a_i,\theta_i)\in \{\0,\1\} \times \Theta_i $ and $\gamma \in \{\0,\1\}$, we have
\begin{itemize}
  \item 
  If $\theta_i = \emptyset$, then 
  \begin{equation}\label{eq:MD-1}
  \texttt{Move}_i^D(a_i,\theta_i, \gamma)
  = \{   (\gamma,   \emptyset), (a_i,  \{(\gamma, N_i^D)\} )     \}
  \end{equation}
  \item 
  If $\theta_i \neq \emptyset$  and $n_{min}=1$, then 
  \begin{equation}\label{eq:MD-2}
  \texttt{Move}_i^D(a_i,\theta_i, \gamma)
  = \{   (\gamma_{min} ,  \textsc{NX}(\theta_i)\cup \{(\gamma, N_i^D )\}) \}
  \end{equation}
  \item 
  If $\theta_i \neq \emptyset$  and $n_{min}>1$, then 
  \begin{align}\label{eq:MD-3}
    &\texttt{Move}_i^D(a_i,\theta_i, \gamma) \\
  = &\left\{ \!\! 
  \begin{array}{c c}
    (a_i  ,  \textsc{NX}(\theta_i)\cup \{(\gamma, N_i^D )\}), \\
    (\gamma_{min}  ,  \textsc{NX}(\theta_i \setminus \{(\gamma_{min},n_{min})  \}  )\cup \{(\gamma, N_i^D )\})
  \end{array} \!\!
  \right\} \nonumber
  \end{align} 
  where $(\gamma_{min},n_{min}) \in \theta$ 	is the pair such that 
  $n_{min} =\min \{n_1,\dots, n_k\}$.
\end{itemize}

Intuitively, $\texttt{Move}_i^D$ captures how the configurations of $\Channel_i$ and $\Act_i$ evolve when   a new decision is issued.  
Note that $\texttt{Move}_i^D$ is a \emph{non-deterministic} transition function as control delay may happen or not. 
More precisely:
\begin{itemize}
\item  
Equation~\eqref{eq:MD-1} captures the case when $\Channel_i$ is empty. 
Depending on whether or not control delay occurs, there are two scenarios
\begin{itemize}
	\item 
	Decision $\gamma$ is received by $\Act_i$ without delay: 
	then  the  configuration of $\Act_i$ will be updated to $\gamma$ and the configuration of $\Channel_i$ is still $\emptyset$; 
	\item 
	Decision $\gamma$ is delayed in $\Channel_i$:  
	then the  configuration of $\Act_i$ will be unchanged while the configuration of $\Channel_i$ will be updated to $\{ (\gamma, N_i^D) \}$, where $N_i^D$ is the upper bound for delays.
\end{itemize} 
\item 
Equation~\eqref{eq:MD-2} captures the case when $\Channel_i$ is full in the sense that there exists a control decision $\gamma_{min}$ with only one unit left, i.e., $n_{min}=1$. 
Since we consider FIFO channel, $\Act_i$ will receive $\gamma_{min}$ definitely 
and the remaining times of all other decisions in $\Channel_i$ will be decreased by one unit with new decision  $(\gamma, N_i^D)$ plugged in. 
\item 
Equation~\eqref{eq:MD-3} captures the case when $\Channel_i$ is non-empty but is also not full. Then the following two scenarios are possible
\begin{itemize}
	\item 
	No decision is received by  $\Act_i$. 
	In this case, the actuator configuration remains unchanged and the channel configuration is updated to $\textsc{NX}(\theta_i)\cup \{(\gamma, N_i^D )\}$. 
	\item 
	Decision $\gamma_{min}$ is received by $\Act_i$. 
	In this case, the actuator configuration is updated to $\gamma_{min}$ and the channel configuration is updated by eliminating $\gamma_{min}$ and adding $\gamma$. 
	Still, no decision whose remaining time is greater than $n_{min}$ can be received due to the FIFO property of the channel.
\end{itemize} 
\end{itemize} 

\begin{remark}
Note that, in $\Theta_i$, the domain of each $(\gamma,n)\in \Theta_i$ is $\{\0,\1\}\times [1,N_i^D]$. 
Therefore, when $N_i^D=0$, i.e., $\Channel_i$ is not suffering from control delay,   the channel configuration is always empty, i.e.,  $\theta_i=\emptyset$. 
In this case, operator $\texttt{Move}_i^D$ is always in the form of
 \[
 \texttt{Move}_i^D (a_i, \emptyset, \gamma)  = (\gamma, \emptyset),\forall a_i,\gamma \in \{\0,\1\}.
 \]
This recovers the standard non-networked setting, where we do not need to specify the actuator configuration separately because it is always the case that the actuator configuration is the same as the latest control decision made by the supervisor. 
However, separating the roles of $\Act_i$ and $\Sup_i$ is crucial in our networked setting. 
\end{remark}
 
We illustrate the above concepts for the case of communication delays by the following example.
 
\begin{figure}
	\centering
	\includegraphics[scale=0.6]{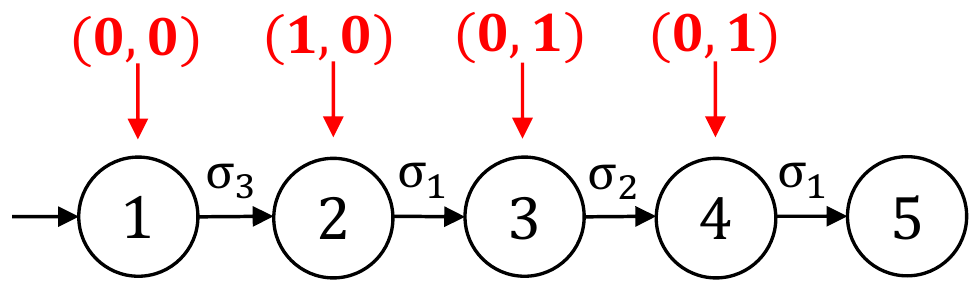}
	\caption{System $\mathbf{G}$ with $\Sigma = \{\sigma_1, \sigma_2, \sigma_3\}$ and  $\Sigma_{uc} = \{\sigma_3\}$.}\label{system}\vspace{-10pt}
\end{figure}

\begin{example}\label{ex:1}
Consider  system $\mathbf{G}$ shown in Figure~\ref{system} with $\Sigma = \{\sigma_1, \sigma_2, \sigma_3\}, \Sigma_c = \{\sigma_1, \sigma_2\}$.  
Suppose that  $\mathbf{G}$ is controlled by a given supervisor $\Sup$. 
Since we assume that the system is fully observed, the vector associated with each state denotes the control decision of $\Sup$ for each $\sigma_i$. 
For example, $(\0,\0)$ at state $1$ represents that the initial control decisions are $\Sup_1(\epsilon)=\0$ and $\Sup_2(\epsilon)=\0$.

Now, let us consider how $\Channel_2$ for event $\sigma_2$ evolves under communication delays. Here we assume $N_2^D = 2$. 
Initially, the configuration for $\Act_2$ is $a_2=\Sup_2(\epsilon)=\0$ 
and the configuration for $\Channel_2$ is $\theta_2=\emptyset$. 
When new event $\sigma_3$ occur, the supervisor sends $\gamma=\Sup_2(\sigma_3)=\0$ to $\Channel_2$ and we update the  configurations to 
\[
\texttt{Move}_2^D (  \0, \emptyset, \0     ) 
= \{ (\0, \emptyset),  (\0, \{ (\0, 2) \})   \}, 
\]
where $(a_2,\theta_2)=(\0, \emptyset)$ corresponds to the case that the newly issued $\0$ is received by $\Act_2$ and 
$(a_2,\theta_2)= (\0, \{ (\0, 2) \}) $ corresponds to the case that the newly issued $\0$ is delayed in $\Channel_2$ and $\Act_2$ is still using the original disable decision $\0$. 
Let us assume that the second case happens and the current configurations are $(a_2,\theta_2)= (\0, \{ (\0, 2) \}) $. 
Then when event $\sigma_1$ occurs,  the supervisor will send $\gamma=\Sup_2(\sigma_3\sigma_1)=\1$ to $\Channel_2$ and then we update the  configurations to 
\[
 \texttt{Move}_2^D (  \0, \{ (\0, 2) \}, \1  ) = 
\{  (\0, \{ (\1,2) \}) ,  (\0,  \{ (\0,1), (\1,2)\}) \}, 
\]
where 
(i) 
$(a_2,\theta_2)=(  \0, \{ (\1, 2) \} )$ corresponds to the case that the previous issued decision $\0$ in the channel is received by $\Act_2$ and the newly issued decision $\1$ is waiting in the channel; and 
(ii) 
$(a_2,\theta_2)= (\0,  \{ (\0,1), (\1,2)\}) $ corresponds to the case that both the previous issued $\0$ and the newly issued $\1$ are delayed in $\Channel_2$ and $\Act_2$ is still using the initial decision $\0$.
\end{example}

\subsection{Communication Losses}  

Now we proceed to model communication losses in the control channel.  
As we mentioned earlier, we assume that communication losses can happen only when the supervisor sends a control decision to a communication channel. 
In other words, once a decision goes into the communication channel, it will not be lost in between and will arrive at the actuator eventually. 

To focus on the effect of control losses, we first do not consider control delays in this subsection. Therefore, we do not need the channel configuration as a part of the "state". However, to capture the assumption  that $\Channel_i$ cannot have more than $N_i^L$ consecutive losses, we need to introduce a ``counter" component.  
We denote by $\mathbb{N}_i^L := \{0,1,\cdots,N_i^L\}$ the counter set.

Now, for controllable event $\sigma_i\in \Sigma_c$, 
suppose that the current actuator configuration is $a_i\in \{\0,\1\}$ and the counter number is $c_i  \in \mathbb{N}_i^L$. 
Suppose that $\Sup_i$ sends a new decision $\gamma \in \{\0,\1\}$. 
We  are also interested in what are the next actuator configuration and the next counter number for $\sigma_i$. 
To this end, we define operator 
\[
\texttt{Move}_i^L : \{\0,\1\}\times \mathbb{N}_i^L \times \{\0,\1\}   \to 2^{ \{\0,\1\}\times  \mathbb{N}_i^L }
\]
by: for any  $a_i\in \{\0,\1\}, c_i\in \mathbb{N}_i^L $ and $\gamma \in \{\0,\1\}$, we have
\begin{itemize}
	\item 
	If $c_i = N_i^L$, then 
	\begin{equation}\label{eq:ML-1}
	\texttt{Move}_i^L(a_i,c_i, \gamma)
	= \{   (\gamma,   0)    \}.
	\end{equation}
	\item  
	If $c_i < N_i^L$, then 
	\begin{equation}\label{eq:ML-2}
	\texttt{Move}_i^L(a_i,c_i, \gamma)
	= \{   (\gamma,   0)  , (a_i,   c_i+1)  \}.
	\end{equation}
\end{itemize}

Intuitively, Equation~\eqref{eq:ML-1} captures the case that there are already $N_i^L$ consecutive losses for control decisions transmissions. 
Therefore, the current decision $\gamma$ will be received by $\Act_i$ definitely and the counter is reset to $0$. 
On the other hand, Equation~\eqref{eq:ML-2} captures the case that the accumulated consecutive losses $c_i$ has not yet exceeded the upper bound. 
Then decision $\gamma$ may either be received, which leads to $(\gamma,0)$, or be lost, which leads to $(a_i,c_i+1)$. 
Therefore, $\texttt{Move}_i^L$ is still a non-deterministic  function in general.

\begin{remark}
When $N_i^L=0$, the above model also boils down to the standard non-networked case. 
In this case, $c_i$ is always equal to $0$, and  $\texttt{Move}_i^L$ is only in the form of 
$\texttt{Move}_i^L(a_i, 0, \gamma) = (\gamma, 0)$, i.e., the control decision can always be received by the actuator immediately. 
\end{remark}

We illustrate the above concepts for the case of communication losses by the following example.

\begin{example}\label{ex:2}
We still consider  system $\mathbf{G}$ shown in Figure~\ref{system} with $\Sigma = \{\sigma_1, \sigma_2, \sigma_3\}, \Sigma_c = \{\sigma_1, \sigma_2\}$.
We assume that   $\Channel_1$ for $\sigma_1$ is subject to communication losses with $N_1^L = 1$ and we illustrate how $\Channel_1$ evolves. 
Initially, the configuration for $\Act_1$ is $a_1=\Sup_1(\epsilon)=\0$ 
and the counter is $c_1=0$. 
When the supervisor sends the first control decision $\gamma=\Sup_1(\sigma_3)=\1$ to $\Channel_1$ upon the occurrence of $\sigma_3$, we update it according to

\[
\texttt{Move}_1^L (  \0, 0, \1     ) 
= \{ (\0, 1),  (\1,  0 )   \}, 
\]
where 
(i) $(\0, 1)$  corresponds to the case that the newly issued decision $\1$ is lost and therefore, $\Act_1$ is still using the previous $\0$ and the countered is added to $1$; and 
(ii) $(\1, 0)$  corresponds to the case that the newly issued decision $\1$ is received by  $\Act_1$ and the counter is reset to $0$.
Note that if the first case happen, 
then the second control decision $\gamma=\Sup_1(\sigma_3\sigma_1)=\0$ to $\Channel_1$ will not be lost for sure as the counter is already $1=N_2^L$.
\end{example}

\subsection{General Case with Both Delays and Losses}
Now, we consider the general case in which both delays and losses may happen in $\Channel_i$. 
In this case, we need to track the channel configuration $\theta_i\in \Theta_i$, the actuator configuration $a_i\in \Act_i$ and the counter number $c_i\in \mathbb{N}_i^L$; these together constitute the "state" of the system, which are also referred to as the \emph{network configuration}. 
To capture how the state evolves upon the occurrence of new decision $\gamma\in \{ \0,\1\}$, 
we define operator
\[
   \texttt{Move}_i^{DL}: \{\0,\1\} \times \Theta_i\times \mathbb{N}_i^L \times \{\0,\1\} \to 2^{\{\0,\1\} \times \Theta_i\times \mathbb{N}_i^L},
\] 
which   combines $\texttt{Move}_i^{D}$ and $\texttt{Move}_i^{L}$. 
Specifically,   for any  $a_i\in \{\0,\1\},\theta_i\in \Theta, c_i\in \mathbb{N}_i^L $ and $\gamma \in \{\0,\1\}$, we have
\begin{itemize}
   \item 
	 If $\theta_i = \emptyset$ and $c_i = N_i^L$, then
	 \begin{equation}\label{eq:G-1}
	\!\!\! \texttt{Move}_i^{DL} (a_i, \theta_i, c_i, \gamma) 
	 =\left\{  
	 \begin{array}{c c}
	 (\gamma, \emptyset, 0),   \\
	 (a_i, \{ (\gamma, N_i^D) \}  , 0) 
	 \end{array}
	 \right\}.  
	 \end{equation}
   \item 
    If $\theta_i = \emptyset$ and $c_i< N_i^L$, then
   	\begin{equation}\label{eq:G-2}
    \!\!\!\texttt{Move}_i^{DL} (a_i, \theta_i, c_i, \gamma) 
    =\left\{  
    \begin{array}{c c}
    (a_i,\emptyset,c_i+1),\\
    (\gamma, \emptyset, 0),   \\
    (a_i, \{ (\gamma, N_i^D) \}  , 0) 
   \end{array}
   \right\}.
   \end{equation}  
   \item 
    If $\theta_i \neq  \emptyset, n_{min}=1$ and $c_i = N_i^L$, then  
    \begin{align}\label{eq:G-3}
    &\texttt{Move}_i^{DL} (a_i, \theta_i, c_i, \gamma)= \\
    &\left\{  
      (\gamma_{min} ,  \textsc{NX}(\theta_i)\cup \{(\gamma, N_i^D )\},0)
    \right\}. \nonumber
    \end{align}   
   \item 
   If $\theta_i \neq  \emptyset, n_{min}=1$ and $c_i < N_i^L$, then  
   \begin{align}\label{eq:G-4}
   &\texttt{Move}_i^{DL} (a_i, \theta_i, c_i, \gamma)= \\
   &\left\{  
   \begin{array}{c c}
   (\gamma_{min} ,  \textsc{NX}(\theta_i)\cup \{(\gamma, N_i^D )\},0),\\
   (\gamma_{min} ,  \textsc{NX}(\theta_i)  ,c_i+1\}) 
    \end{array}
    \right\}.  \nonumber
   \end{align}       
   \item 
   If $\theta_i \neq  \emptyset, n_{min}>1$ and $c_i = N_i^L$, then  
   \begin{align}\label{eq:G-5}
   &\texttt{Move}_i^{DL} (a_i, \theta_i, c_i, \gamma)= \\
   & \left\{ 
   \begin{array}{c c}
   (\gamma_{min}, \textsc{NX}(\theta_{i} \setminus \{ (\gamma_{min}, n_{min}) \} ) \cup \{ (\gamma, N_i^D) \}, 0 ), \\
   (a_i, \textsc{NX} (\theta_i) \cup \{  (\gamma, N_i^D) \}, 0)
   \end{array} 
   \right\}. \nonumber
   \end{align}   
   \item 
   If $\theta_i \neq  \emptyset, n_{min}>1$ and $c_i < N_i^L$, then  
   \begin{align}\label{eq:G-6}
   &\texttt{Move}_i^{DL} (a_i, \theta_i, c_i, \gamma)= \\
   &\left\{
   \begin{array}{c c}
    (a_i, \textsc{NX} (\theta_{i}), c_i + 1), \\
    (\gamma_{min}, \textsc{NX}(\theta_{i} \setminus \{ (\gamma_{min}, n_{min}) \} ), c_i + 1), \\
    (a_i, \textsc{NX} (\theta_i) \cup \{ (\gamma, N_i^D)\} , 0 ),\\
   (\gamma_{min}, \textsc{NX}(\theta_{i} \setminus \{ (\gamma_{min}, n_{min}) \} ) \cup \{ (\gamma, N_i^D) \}, 0 )
   \end{array} 
   \right\}.  \nonumber
   \end{align}       
\end{itemize}

The above defined  $\texttt{Move}_i^{DL}$ boils down to 
$\texttt{Move}_i^{D}$ or $\texttt{Move}_i^{L}$, respectively, when $N_i^L=0$ or $N_i^D=0$. 
Specifically, when $N_i^L=0$, Equations~\eqref{eq:G-1},~\eqref{eq:G-3} and~\eqref{eq:G-5} becomes Equations~\eqref{eq:MD-1},~\eqref{eq:MD-2} and~\eqref{eq:MD-3}, respectively. 
When   $N_i^D=0$, Equations~\eqref{eq:G-1}  and~\eqref{eq:G-2} becomes Equations~\eqref{eq:ML-1}  and~\eqref{eq:ML-2}, respectively. 
However, Equations~\eqref{eq:G-4} and~\eqref{eq:G-6} capture the scenarios when delays and losses are combined. 
Specifically, the first element in Equation~\eqref{eq:G-4} corresponds to the case that $\gamma_{min}$ is received by $\Act_i$ but $\gamma$ is plugged in $\Channel_i$, while the the second element  corresponds to the case that $\gamma_{min}$ is received by $\Act_i$ but $\gamma$ is lost.
For Equation~\eqref{eq:G-6}, it captures the four possible combinations of "$\gamma_{min}$ is received by $\Act_i$ or not" and "$\gamma$ is lost or not".

\section{Closed-loop Dynamics of the Overall System}\label{sec:4}

In the previous section, we have discussed how to model the state and dynamic of each \emph{single} channel. In this section, we aim to consider all (independent) channels together and provide the closed-loop behavior of the overall supervisory control system with control delays and losses.

\subsection{Multiple Channel System}
Since control channels are independent, the overall "state" of the systems is a vector in which each component reflects the status of a single channel. 
Formally, 
the actuator configuration of the overall system is a $|\Sigma_c|$-dimensional vector 
\[
\textbf{a}=(a_1,a_2,\dots,a_{|N_c|})\in \{\0,\1\}^{|\Sigma_c|}.
\] 
We denote by $A$ the set of all actuator configurations.
For each actuator configuration $\textbf{a}\in A$, we denote by $\Gamma(\textbf{a})$ the set of controllable events that are enabled, i.e., 
\[
\Gamma(\textbf{a})=\{ \sigma_i\in \Sigma_c: a_i=\1  \}.
\]
Similarly, the channel configuration of the overall system  is a $|\Sigma_c|$-dimensional vector 
\[
\boldsymbol{\theta}=(\theta_1,\theta_2,\dots,\theta_{|N_c|})\in  \Theta_1\times\Theta_2\times\cdots\times \Theta_{|\Sigma_c|}.
\] 
We denote by $\Theta=\Theta_1\times\cdots\times\Theta_{|\Sigma_c|}$ the set of all channel configurations. 
Also, we define the overall consecutive losses counter as
\[
\textbf{c} = (c_1,c_2,\cdots,c_{|\Sigma_c|})\in \mathbb{N}_1^L\times \mathbb{N}_2^L  \times \cdots\times   \mathbb{N}_{|\Sigma_c|}^L 
=:\mathbb{N}^L
\]
The dynamics of the overall network configuration is defined by operator
\[
\texttt{Move} : A\times \Theta  \times \mathbb{N}^L \times \{\0,\1\}^{|\Sigma_c|} \to 2^{ A\times \Theta \times \mathbb{N}^L}
\]
such that: 
for any  
$( \textbf{a},  \boldsymbol{\theta} ,\textbf{c}), ( \textbf{a}',  \boldsymbol{\theta}' ,\textbf{c}') \in A\times \Theta \times \mathbb{N}^L$ and $\boldsymbol{\gamma} \in \{\0,\1\}^{|\Sigma_c|}$ , we have $(\textbf{a}', \boldsymbol{\theta}', \textbf{c}')\in \texttt{Move}(\textbf{a}, \boldsymbol{\theta}, \textbf{c}, \boldsymbol{\gamma})$ iff
\begin{align}
 \forall i=1,\dots,|\Sigma_c|:(a_i', \theta_i', c_i')\!\in\! \texttt{Move}_i^{DL}(a_i,\theta_i, c_i,\gamma_i). 
\end{align}

\subsection{Closed-loop Language} 
In order to specify the complete dynamics of the closed-loop system, the last piece is to specify the  initial configuration of the network. 
As we mentioned early, the initial control decision can neither be delayed nor lost. 
In practice, this means that the initial decisions are already embedded in the actuators before the plant runs. 
Therefore, the initial actuator configuration is given by 
\[
\textbf{a}_0=( \Sup_{1}(\epsilon),\Sup_{2}(\epsilon),\dots, \Sup_{|\Sigma_c|}(\epsilon)   )\in A.
\]
Since the communication channels are all empty initially, 
the initial channel configuration is given by
\[
\boldsymbol{\theta}_0=(\emptyset, \emptyset, \dots,\emptyset )\in \Theta.
\]
Also, the initial counter is given by 
\[
\textbf{c}_0=( 0,0,\dots, 0 )\in \mathbb{N}^L.
\]

Note that, although our main purpose is to calculate the actual strings generated by the system, to this end, we also need to track the configurations of the overall network simultaneously. 
This is captured by the concept of \emph{extended strings}.
Formally,  an extended string is a string augmented with the network configuration  in the form of 
\[
t=(s, \textbf{a}, \boldsymbol{\theta},\textbf{c}) \in \Sigma^*\times A\times\Theta\times \mathbb{N}^L.
\]
A set of extended string is called an extended language. 
Therefore, a networked supervisory control system essentially generates an extended language
$\mathcal{L}_e(\Sup/\mathbf{G})$ defined recursively as follows:
\begin{itemize}
	\item 
	$(\epsilon, \textbf{a}_0,\boldsymbol{\theta}_0,\textbf{c}_0 ) \in \mathcal{L}_e(\Sup/\mathbf{G})$; 
	\item 
	For any $t=(s, \textbf{a},\boldsymbol{\theta},\textbf{c})$ and $\sigma\in \Sigma$, 
	we have $t'=(s', \textbf{a}',\boldsymbol{\theta}',\textbf{c}' ) \in \mathcal{L}_e(\Sup/\mathbf{G})$ iff 
	\begin{itemize}
		\item 
		$t  \in \mathcal{L}_e(\Sup/\mathbf{G})$; 
		\item 
		$s'= s\sigma \in \mathcal{L}(\mathbf{G}) $; 
		\item 
		$\sigma\in  \Gamma(  \textbf{a}  ) \cup \Sigma_{uc}$; 
		\item 
		$(\textbf{a}',\boldsymbol{\theta}',\textbf{c}') \in \texttt{Move}(  \textbf{a},\boldsymbol{\theta},\textbf{c},  \boldsymbol{\gamma}(s') )$, where $\boldsymbol{\gamma}(s'):=(\Sup_1(s'),\dots, \Sup_{|\Sigma_c|}(s'))  $
	\end{itemize}
\end{itemize}

The above definition is explained as follows. 
Essentially, it defines all extended strings that can be generated by the system in an inductive manner. 
In the inductive step, the first condition says that the prefix should be an extended string generated by the system. 
The second condition says that new event $\sigma$ should be feasible in the original plant $\mathbf{G}$. 
The third condition says that event $\sigma$ should be enabled by the current actuator configuration $\textbf{a}$. 
Finally, the last condition captures how the network configuration evolves upon the new control decision $\boldsymbol{\gamma}(s')$. 
The effect of $\boldsymbol{\gamma}(s')$ will be recorded in the network configuration and may determine the occurrence of some future events when it arrives at actuators as $\textbf{a}$. 
 
Based on the extended language, we can define the actual language generated by the closed-loop system in the networked setting as the projection of the extended language onto the first component, i.e., 
\[
\mathcal{L}(\Sup/\mathbf{G})
=\{s\!\in \!\Sigma^*:  
\exists \textbf{a},\boldsymbol{\theta},\textbf{c}\text{ s.t.}
(s, \textbf{a},\boldsymbol{\theta},\textbf{c} ) \!\in\! \mathcal{L}_e(\Sup/\mathbf{G} )
\}.
\]
This completes our general model for networked supervisory control system over multiple channel networks.

\subsection{Case of Shared Control Channels}
Note that, in all the above developments, it is assumed that each actuator uses its own control channel. 
In some applications, a group of actuators may share the same control channel because they are physically located together and  the control decisions are sent as packages to each group of actuators.
In such a scenario, we can assume that $\Sigma_c$ is partitioned as $m$ groups
\[
\Sigma_c=\Sigma_{c,1}\dot{\cup}\Sigma_{c,2}\dot{\cup}\dots \dot{\cup}\Sigma_{c,m}
\]
such that actuators  in each $\Sigma_{c,i}$ share the same control channel $\Channel_i$. 
Our framework and all previous developments can be easily adapted to the case. 
The only differences is that the control decision in each $\Channel_i$ is then in the domain of $\{\0,\1\}^{|\Sigma_{c,i}|}$ rather than in $\{\0,\1\}$. 
In the most extreme case where all actuators share the same control channel, this general setting boils down to the single channel setting in the framework of Lin \cite{lin2014control}, where all decisions are sent as a single package. 
Here, instead of formally presenting the generalization which is rather straightforward, we use the following example to illustrate our point.	

\begin{example}\label{ex:compare}
We still consider  system $\mathbf{G}$ shown in Figure~\ref{system} with   $\Sigma_c = \{\sigma_1, \sigma_2\}$.
Now assume that both events $\sigma_1$ and $\sigma_2$ share the same control channel, which is the single channel denoted by $\Channel$. 
This actually corresponds to the setting in \cite{lin2014control}, where all control decisions are sent as a package. 
Therefore, each control decision in the channel is in $\{\0,\1\}\times \{\0,\1\}$. 
For instance, $\gamma=(\1,\0)$ at state $2$ means that   $\Sup_{1}(\sigma_3) = \1$ and $\Sup_2(\sigma_3) = \0$.

Now, suppose that  $\Channel$ is subject to communication delays with $N^D = 2$ and communication losses with $N^L = 1$. 
We still consider supervisor $\Sup$ as specified in Figure~\ref{system}. 
Then the extended language generated by $\Sup/\mathbf{G}$ can be computed as follows:
\begin{itemize}
  \item 
   Initially, we have 
   \[
   (\epsilon, \textbf{a}_0, \theta_0, \textbf{c}_0) = (\epsilon, (\0,\0), \emptyset, 0) \in \mathcal{L}_e (\Sup/\mathbf{G}),
   \]
   Note that the dimensions of $\theta_0$ and $\textbf{c}_0$ are both one since there is only one single channel. 
   \item  
   Since $\sigma_3$ is uncontrollable and $\sigma_3\in \mathcal{L} (\mathbf{G})$, 
   we compute 
   \begin{align}
   &\texttt{Move} (\textbf{a}_0, \boldsymbol{\theta}_0, \textbf{c}_0, \gamma )\nonumber
   =\\
   &\{
   (  (\0,\0), \{( (\1,\0), 2)\}, 0), 
   (  (\0,\0), \emptyset,       1),
   (  (\1,\0), \emptyset,       0)
   \}\nonumber
   \end{align}
   with $\gamma = (\Sup_1(\sigma_3), \Sup_2(\sigma_3))=(\1,\0)$. 
   Therefore, the following three extended strings are in $\mathcal{L}_e (\Sup/\mathbf{G})$:
	\begin{itemize}
			\item $(\sigma_3, (\0,\0), \{(  (\1,\0), 2)\}, 0)$; and 
			\item $(\sigma_3, (\0,\0), \emptyset, 1)$; and
			\item $(\sigma_3, (\1,\0), \emptyset, 0)$.
	\end{itemize}
	which represent the cases that the newly issued control decision (package) $\gamma=(\1,\0)$ is delayed, lost and neither delayed nor lost, respectively.
	\item 
	When the actuator configuration is $(\0,\0)$, we have $\Gamma((\0,\0)) = \emptyset$, which means both $\sigma_1$ and $\sigma_2$ are disabled; therefore, no event can happen from state $2$. 
	However, for $\Gamma( (\1,\0) ) = \{\sigma_1\}$, we can  enable $\sigma_1$ for extended string $(\sigma_3, (\1,\0), \emptyset, 0)$. 
	Then we compute 
	\begin{align}
	&\texttt{Move} ((\1,\0), \emptyset, 0, \gamma )\nonumber
	=\\
	&\{
	(  (\1,\0), \{( (\0,\1), 2)\}, 0), 
	(  (\1,\0), \emptyset,       1),
	(  (\0,\1), \emptyset,       0)
	\}\nonumber
	\end{align}
	with $\gamma = (\Sup_1(\sigma_3\sigma_1), \Sup_2(\sigma_3\sigma_1))=(\0,\1)$. 
	This yields the following three extended strings  in $\mathcal{L}_e (\Sup/\mathbf{G})$:  
	     \begin{itemize}
	     	\item 
	     	$(\sigma_3\sigma_1, (\1,\0), \{((\0,\1),2)\}, 0)$; and 
	     	\item 
	     	$(\sigma_3\sigma_1, (\1,\0), \emptyset, 1)$; and 
	     	\item 
	     	$(\sigma_3\sigma_1, (\0,\1), \emptyset, 0)$.
	     \end{itemize}
     \item 
     For the last extended string above, we have 
     $\Gamma( (\0,\1) ) = \{\sigma_2\}$, i.e., event $\sigma_2$ is enabled.
     By computing $\texttt{Move} ((\0,\1), \emptyset, 0, \gamma)$ 
     for $\gamma = (\Sup_1(\sigma_3\sigma_1\sigma_2), \Sup_2(\sigma_3\sigma_1\sigma_2))=(\0,\1)$, we obtain the following three extended strings:
         \begin{itemize}
         	\item $(\sigma_3\sigma_1\sigma_2, (\0,\1),\{((\0,\1),2)\}, 0)$; and 
         	\item $(\sigma_3\sigma_1\sigma_2, (\0,\1), \emptyset, 1)$; and 
         	\item $(\sigma_3\sigma_1\sigma_2, (\0,\1),\emptyset,0)$.
         \end{itemize}
     \item 
     Finally, since
     $\Gamma( (\0,\1) ) = \{\sigma_2\}$ for each of the above extended string, 
     we can no longer proceed to enable $\sigma_1$, i.e., no extended string can be defined from $\sigma_3\sigma_1\sigma_2$.    
	\end{itemize} 
	Therefore, by projecting  the obtained extended strings above to their first components, we have $\mathcal{L}(\Sup/\mathbf{G}) = \{\epsilon, \sigma_3,\sigma_3\sigma_1,\sigma_3\sigma_1\sigma_2\}$. 
\end{example}

\begin{remark}
In the proposed modeling framework, the state-space grows exponentially fast with the delay upper bound $N^D_i$ and the loss upper bound $N^L_i$ due to the combination of configurations of control channels. 
However, in practice, $N^D_i$ and $N^L_i$ are usually very small and delays/losses occur very rarely.  For the case of large delays and losses, one may first seek to improve the network hardware environment rather than to improve the control algorithm.
\end{remark}

\section{Comparison and Discussion}\label{sec:5}
In \cite{lin2014control}, Lin proposed a language-based framework for supervisory control of networked discrete event systems.
In this section, we compare the our model for networked supervisory control systems with the one proposed in \cite{lin2014control} and show that our model more precisely captures the practical situation compared with Lin's model. 

Note that the framework in \cite{lin2014control} considers delays and losses in both control and observation channels. For the purpose of comparison, here we just review the part for delays and losses in control channels. 
The basic setting of control delays and losses in \cite{lin2014control} is as follows:
\begin{itemize}
	\item 
	The control decision made by the supervisor is sent as a package to the actuators; and 
	\item 
	Communication delays in control are assumed to be bounded by $N_c$ steps and at least one control command in the past $N_c$ steps can be  received by the supervisor.
\end{itemize}
Essentially, the first setting assumes that a single control channel	is used in the feedback loop. However, our framework captures the more general setting with multiple control channels. 
For the second setting,  \cite{lin2014control} proposes the following definition of closed-loop language $L(\Sup/\mathbf{G})$ \footnote{We used $L(\cdot)$ rather than $\mathcal{L}(\cdot)$ to distinguish between the closed-loop language defined in \cite{lin2014control} with our definition.}:
\begin{itemize}
	\item 
	$\epsilon\in L(\Sup/\mathbf{G})$; 
	\item 
	for any $s\in \Sigma^*,\sigma\in \Sigma$, we have $s\sigma \in L(\Sup/\mathbf{G})$ iff
	\begin{itemize}
		\item
		$s\in L(\Sup/\mathbf{G})$; and 
		\item 
		$s\sigma\in L(\mathbf{G})$; and 
		\item
		$\sigma \!\in\! \Sigma_{uc}\!\cup\! \Sup(s)\!\cup\! \Sup(s_{-1})\!\cup\!\cdots \!\cup\! \Sup(s_{-N_c})$
	\end{itemize}
\end{itemize}
where $s_{-k}$ denotes the string obtained by removing the last $k$ events in $s$. 
This definition essentially handles control delays by roughly considering all possible control decision in the past $N_c$ steps. 
Since all possible control decisions are considered, control losses are handled implicitly as the case of no decision received is also included. 

Therefore, such a definition does not precisely distinguish between control delays and losses precisely as both are handled roughly. As a consequence, $L(\Sup/\mathbf{G})$ is just  an \emph{over-approximation} of the behavior of the closed-loop system and it may contain some behaviors that are physically not possible. 
We illustrate this point by the following example. 
\begin{example}
We consider system $G$ shown in Figure~\ref{system} with 
$\Sigma_c = \{\sigma_1, \sigma_2\}$.
We assume there is only one single control channel, i.e., $\sigma_1$ and $\sigma_2$ shares the same channel as the case of  Example~\ref{ex:compare}. 
Also, we assume the delay bound is $N^D=2$ or  $N_c = 2$ using the notation of \cite{lin2014control}. 
  
According to the definition of closed-loop language in \cite{lin2014control}, 
we can compute $L(\Sup/\mathbf{G})$ recursively by:
  \begin{itemize}
  	 \item $\epsilon \in L(\Sup/\mathbf{G})$; 
  	 \item $\sigma_3\in L(\Sup/\mathbf{G})$ since $\sigma_3 \in \Sigma_{uc}$
  	 \item $\sigma_3\sigma_1\in L(\Sup/\mathbf{G})$ since $\sigma_1\in \Sup(\sigma_3)$;
  	 \item $\sigma_3\sigma_1\sigma_2\in L(\Sup/\mathbf{G})$ since $\sigma_2\in \Sup(\sigma_3\sigma_1)$;
  	 \item $\sigma_3\sigma_1\sigma_2\sigma_1\in L(\Sup/\mathbf{G})$ since $\sigma_1\in \Sup((\sigma_3\sigma_1\sigma_2)_{-2})$. 
  \end{itemize}
However, the last string $\sigma_3\sigma_1\sigma_2\sigma_1$ is not physically possible 
because the enablement of the last $\sigma_1$ relies on decision $\Sup((\sigma_3\sigma_1\sigma_2)_{-2})=\Sup(\sigma_3)$ which is received at the instant of $\sigma_3$. 
However, 
the occurrence of $\sigma_3\sigma_1\sigma_2$ means that the actuator must have received 
$\Sup(\sigma_3\sigma_1)=\sigma_2$ without delay or loss. 
This means that $\Sup(\sigma_3)$ has already been (i) taken from the channel, 
(ii) received by the actuator; and (iii) erased by new decision received by the actuator. 
Therefore, event $\sigma_1$ after $\sigma_3\sigma_1\sigma_2$ is not possible physically. 
However, this string is included in the definition of \cite{lin2014control} because it only considers an estimated possible control decisions, in which some may not be feasible.  

However, as we have already discussed in Example~\ref{ex:compare} (assuming there is no control loss), 
string $\sigma_3\sigma_1\sigma_2\sigma_1$ is not included in $\mathcal{L}(\Sup/\mathbf{G})$ because we precisely model the state-space and the dynamics of the network. 
This example also justifies the advantage of the proposed network model compared with the existing one in the literature in addition to the general multiple channel setting.
\end{example}

\section{Conclusion}
In this paper, we provide a new framework for  modeling and analysis of networked supervisory control systems with control delays and control losses over multiple communication channel networks. 
Our approach follow a system theoretic perspective by identifying the state-space of overall networked system and investigating the dynamic of the entire state-space. 
Compared with the language-based definition of networked DES, our framework more precisely captures the behavior of a networked DES that are physically feasible. Furthermore, our framework allow to handle the general scenario where the supervisor may send control decisions to different actuators via different communication channels.

Note that, throughout this paper, we investigate  how the networked system evolves under a \emph{given} supervisor and define its closed-loop language appropriately. Two immediate problems arise within this framework: 
(i) whether or not there exists a supervisor achieving a given specification language; and 
(ii) how can we synthesize a supervisor whose closed-loop behavior is a sub-language of the specification. 
The former is the supervisor existence problem and the latter is the supervisor synthesis problem. We plan to investigate them within our framework in the future. Also, we would like to consider the issue of delays and losses in observation channels together with the control channels.

\bibliographystyle{plain}
\bibliography{reference}

\end{document}